\pdfoutput=1

\documentclass[conference]{IEEEtran}
\IEEEoverridecommandlockouts

\usepackage{cite}
\usepackage{amsmath,amssymb,amsfonts}
\usepackage{algorithmic}
\usepackage{graphicx}
\usepackage{textcomp}
\usepackage{xcolor}

\usepackage{hyperref}
\usepackage{enumitem}

\newtheorem{definition}{Definition}[section]

\def\BibTeX{{\rm B\kern-.05em{\sc i\kern-.025em b}\kern-.08em
    T\kern-.1667em\lower.7ex\hbox{E}\kern-.125emX}}
    
\begin{document}

\title{CODY: A graph-based framework for the analysis of COnversation DYnamics in\\online social networks

    \thanks{We thank the Klaus Tschira Foundation for funding this research in the framework of the EPINetz project: \url{https://epinetz.de}.}
}

\author{\IEEEauthorblockN{1\textsuperscript{st} John Ziegler}
    \IEEEauthorblockA{\textit{Institute of Computer Science} \\
        \textit{Heidelberg University}\\
        Heidelberg, Germany \\
        ziegler@informatik.uni-heidelberg.de}
    \and
    \IEEEauthorblockN{2\textsuperscript{nd} Fabian Kneissl}
    \IEEEauthorblockA{\textit{Institute of Computer Science} \\
        \textit{Heidelberg University}\\
        Heidelberg, Germany \\
        fabian.kneissl@stud.uni-heidelberg.de}
    \and
    \IEEEauthorblockN{3\textsuperscript{rd} Michael	Gertz}
    \IEEEauthorblockA{\textit{Institute of Computer Science} \\
        \textit{Heidelberg University}\\
        Heidelberg, Germany \\
        gertz@informatik.uni-heidelberg.de}
}

\maketitle

\begin{abstract}
    Conversations are an integral part of online social media, and gaining insights into these conversations is of significant value for many commercial as well as academic use cases. From a computational perspective, however, analyzing conversation data is complex, and numerous aspects must be considered. Next to the structure of conversations, the discussed content -- as well as their dynamics -- have to be taken into account. Still, most existing modelling and analysis approaches focus only on one of these aspects and, in particular, lack the capability to investigate the temporal evolution of a conversation. To address these shortcomings, in this work, we present CODY, a content-aware, graph-based framework to study the dynamics of online conversations along multiple dimensions. Its capabilities are extensively demonstrated by conducting three experiments based on a large conversation dataset from the German political Twittersphere. First, the posting activity across the lifetime of conversations is examined. We find that posting activity follows an exponential saturation pattern. Based on this activity model, we develop a volume-based sampling method to study conversation dynamics using temporal network snapshots. In a second experiment, we focus on the evolution of a conversation's structure and leverage a novel metric, the temporal Wiener index, for that. Results indicate that as conversations progress, a conversation's structure tends to be less sprawling and more centered around the original seed post. Furthermore, focusing on the dynamics of content in conversations, the evolution of hashtag usage within conversations is studied. Initially used hashtags do not necessarily keep their dominant prevalence throughout the lifetime of a conversation. Instead, various "hashtag hijacking" scenarios are found. Also, during the conduction of this work, we developed a software package to facilitate the analysis of heterogeneous information networks. It is made publicly available to be used by the research community.
\end{abstract}


\begin{figure*}[t]
    \centering
    \includegraphics[width=\textwidth]{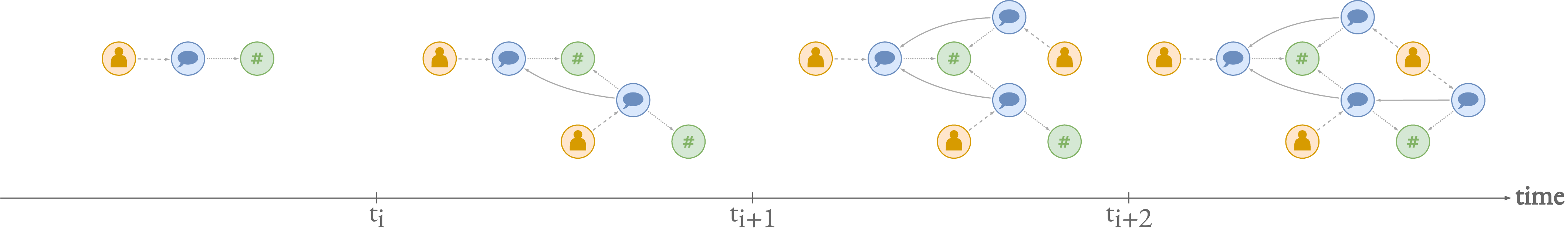}
    \caption{The graph-based CODY model incorporates three central dimensions of a conversation: structure, content and dynamics. Different node and edge types are represented by different symbols and line styles, respectively. Temporal snapshots enable the dynamics of the conversation to be shown.}
    \label{fig:cody_model}
\end{figure*}

\section{Introduction}\label{sec:introduction}

Interpersonal communication is integral to many social media platforms and is well-known through conversational features such as Twitter replies or YouTube comments. Extracting valuable information from these conversations can be beneficial for many commercial as well as academic use cases. For this, elaborate modelling and analysis approaches are required. Still, existing approaches show several deficiencies. In a recent work by Brambilla et al. \cite{brambilla2022graph}, the authors state that "[m]ost studies on social networks have focused only on user relationships \emph{or} [emphasis added] on the shared content, while ignoring the valuable information hidden in the digital conversations, in terms of structure of the discussion and relation between contents, which is essential for understanding online communication behavior." They highlight the need for an analysis model to incorporate a conversation's content \emph{and} structure. While this insight leads towards a more holistic conversation analysis model, it still lacks an essential aspect. As conversations are fundamentally characterized by their temporal evolution, a respective analysis model should also consider a conversation's dynamics. To the best of our knowledge, existing conversation analysis approaches are still missing this aspect \cite{brambilla2022graph, saveski2021structure, cogan2012reconstruction}. We overcome these shortcomings by proposing a graph-based conversation analysis model, named CODY, which incorporates the conversation's \emph{structure}, \emph{content} and \emph{dynamics}. Its capabilities are demonstrated by focusing on three main research questions:

\begin{enumerate}[leftmargin=*, label=RQ\arabic*]
    \item How does the posting activity change over the lifetime of a conversation? Does a generalizing model describe the observed activity pattern?\label{item:rq1}
    \item How do conversations structurally evolve? Is there a metric to quantitatively describe the structural evolution?\label{item:rq2}
    \item How does the focus of discussed topics change over the lifetime of a conversation? Do initially used hashtags keep their dominant prevalence?\label{item:rq3}
\end{enumerate}

To answer these research questions, an analysis model needs to be developed that is versatile enough to cope with the three dimensions of a conversation -- structure, content and dynamics -- as no current approach exists for that. The proposed CODY model is based on the concept of temporal heterogeneous information networks (THINs) (cf. \cite{sun2010community, li2018temporal, milani2019relationship}) and therefore comes with proper flexibility to model a conversation's evolution, along with its structure and content. By leveraging a large conversation dataset from the German political Twittersphere, the model's applicability is shown, and proposed research questions are investigated. The developed methodology and the results of our experiments make up several contributions:

\begin{itemize}
    \item The graph-based CODY model constitutes a versatile framework to study the structure, content and dynamics of conversations. Its flexibility in terms of integrated node/edge types allows to model various (content) objects and their interplay within a conversation.
    \item \ref{item:rq1}: Based on empirical results, the posting activity of a conversation follows an exponential saturation pattern. Contributions occur mainly at the early stage of a conversation, and participation becomes less towards the end.
    \item The developed activity model lays the theoretical foundation for a volume-based sampling approach used to model a conversation's evolution based on temporal snapshots. The common issue of finding an appropriate sampling technique is overcome with that.
    \item \ref{item:rq2}: The newly proposed temporal Wiener index metric allows to quantitatively measure a conversation's structural evolution. Experimental results show that as conversations progress, their structure tends to be less sprawling and more centered around the original post.
    \item \ref{item:rq3}: Focused on the content of conversations, investigations regarding the temporal prevalence of hashtags show that initially used hashtags do not necessarily keep their dominant role across a conversation's lifetime. Instead, various "hashtag hijacking" scenarios can be observed.
    \item Along with the conducted experiments, we developed a software package to facilitate the analysis of heterogeneous information networks. Its source code\footnote{GitHub: \url{https://github.com/codingfabi/hetpy}} is publicly available to be used by the research community.
\end{itemize}

Regarding the used terminology, in line with previous work \cite{cogan2012reconstruction}, we focus on conversations that start from an initial social media post and that are made up of direct user interactions related to this post rather than "conversations" that are centered around an event or other social media entities, such as @mentions. In our understanding, conversations must be based on direct interpersonal communication instead of latent social interactions. In the context of the Twitter platform, replies are considered "[\ldots] the most direct communication sign between two users" \cite{cogan2012reconstruction}. Therefore, Twitter reply threads, as leveraged for the conducted experiments, make up conversations.
\section{Related Work}\label{sec:related_work}

Most closely related to our work are studies about the modelling and analysis of online conversations. As such, Brambilla et al. \cite{brambilla2022graph} conduct an in-depth study of various online conversations. For their analysis, they propose a graph-based framework and mainly focus on intention analysis as well as network construction. In contrast to us, they do not consider the dynamics of the conversation graph itself. Still, they consider temporal aspects of the analyzed conversations, such as their duration and typical reply times. Similar to our approach, they also view the network as heterogeneous by including different node and edge types. Recently, Botzer and Weninger \cite{botzer2023entity} leverage entity graphs extracted from online conversation data, in their case Reddit, to study whether online discourse is predictable, how the online conversations are structured and whether theories on spreading activation can be successfully applied in the context of online discourse analysis. Compared to our work, they do not investigate the conversation graphs' dynamics or consider the heterogeneity of the conversations' content. Further, in one of the early works on conversation graph analysis, Cogan et al. \cite{cogan2012reconstruction} investigate the characteristics of conversation networks emerging from an initial social media seed post. This understanding of a conversation graph is in line with our definition. In their work, they propose a method to gather conversation data as completely as possible and conduct an experimental analysis based on Twitter data. Their outlook highlights the importance of future work considering a conversation graph's temporal evolution. Focused on the aspect of toxicity, Saveski et al. \cite{saveski2021structure} also study the structure of Twitter conversations. They conduct their analysis on different levels, from individual users to the entire conversation group. Even though applied in a static setting, they also leverage the Wiener index to investigate the topology of analyzed reply trees. In two prediction tasks, they aim at forecasting future toxic conversation characteristics based on previous conversation data. In another application-focused work, Mathew et al. \cite{mathew2016mining} analyze Twitter conversation data related to e-commerce promotional events. They model the data as a user-centric network by considering replies, mentions, and retweets for the weighted edges and Twitter users as nodes. Regarding the network's temporal evolution, they determine the users' flow between different network components across time. Unfortunately, they do not elaborate on their temporal graph model but only propose a sampling approach using fixed time windows. Also, they do not investigate the dynamics of any network metric and solely focus on users, not the content, as part of a conversation.

Information diffusion is a topic that is also extensively studied in the context of online social networks. The survey by Guille et at. \cite{guille2013information} and the more recent one by Yujie \cite{yujie2020survey} give a good overview of this research field. Commonly used information diffusion models also leverage topological information of the underlying social network. The spreading process is frequently modelled as a network, referred to as diffusion graph. In contrast to our work, information diffusion mainly investigates how information spreads across the entire network. We, however, are less interested in the information diffusion process but instead focus on the graph-centric evolution of individual online conversations, which includes not only the discussed information but also the participating actors and the structure of a conversation. Still, based on our model, temporal meta paths related to the concept of THINs could be leveraged to study information diffusion processes within individual conversations.

As part of our methodology, we leverage THINs to model the evolution of online conversations. Even though, to the best of our knowledge, none of the existing work applied THINs in the context of conversation analysis, they are extensively used for other use cases. For example, Sun et al. \cite{sun2010community} propose a method to detect communities in THINs. They understand THINs as a sequence of temporal network snapshots, which aligns with our approach. Similarly, Cuzzocrea and Folino \cite{cuzzocrea2013community} deal with community detection in temporal information networks. They specifically focus on the temporal tracking of the detected communities and propose a detection method to analyze structural changes regarding the community structure of the evolving network. Further, again in the context of THINs, Fard et al. \cite{milani2019relationship} tackle the task of meta path prediction. They leverage temporal snapshots to model a network's evolution. Sajadmanesh et al. \cite{sajadmanesh2019continuous} also predict relationships in THINs. Additionally, they predict the time when these relationships will occur. Further, multiple other use cases are elaborated in the context of THINs. These include the engineering of node embeddings \cite{wang2020dynamic} \cite{bian2019network}, sequential recommendation \cite{xie2021sequential}, network motifs \cite{li2018temporal}, and link inference \cite{jia2017link} \cite{aggarwal2012dynamic}.
\section{CODY Model}\label{sec:cody_model}

As already described, a conversation's structure, content and dynamics must be considered for a conversation analysis model to be leverageable in different analysis scenarios. Following the same pattern, this section introduces the CODY model in three steps. First, Section \ref{sec:structural_conversation_model} explains how the CODY model considers the structure of conversations. Secondly, the temporal evolution of a conversation is integrated into the model in Section \ref{sec:temporal_conversation_model}. Finally, Section \ref{sec:content_aware_conversation_model} complements the model by additionally considering the content of conversations. Apart from the conversation model, metrics to investigate a conversation's structural evolution are needed. Therefore, Section \ref{sec:temporal_wiener_index} introduces a novel metric, denoted as \emph{temporal} Wiener index, to quantitatively study how a conversation's structure evolves over time.

\subsection{Structural conversation model}\label{sec:structural_conversation_model}

Considering the structure of a conversation is an essential requirement for a holistic analysis model. On an abstract level, an online conversation consists of posts (e.g., tweets or comments) and links between these posts that represent semantic relationships (e.g., replying or commenting). The concept of graphs\footnote{Both terms,  \emph{graph} and \emph{network}, are used interchangeably in this work. The same is true for the terms "object", "vertex", and "node", as well as "edge", "link", and "relationship".} is particularly suitable for modelling objects, in our case posts, and their relationships. We resort to these graphs as the primary modelling approach for the CODY framework. Posts contained in the conversation are represented as nodes and are linked via semantic relationships represented as edges. For a conversation graph with nodes $V$ and edges $E$ to take the different semantic types into account, the network definition comes with a node type mapping $\phi: V \rightarrow A$ and a link type mapping $\psi: E \rightarrow R$ with $A$ and $R$ denoting the network's node and edge types respectively. This network typing aligns with the concept of HINs (cf. Sun and Han \cite{sun2013mining}).

\begin{definition}[Structural conversation network]\label{def:structural_conversation_network}
    The structural network $G_{struc} = (V, E)$ of a conversation consists of a set of posts modelled as nodes $V$ ($\phi(v) = \text{post}$ with $v \in V$) and a set of links $E \subset V \times V$ among these posts. Each link $e = (v_i, v_j)$ is a tuple of the two adjacent posts $v_i$ and $v_j$.
\end{definition}

A single conversation is treated as an individual network based on Definition \ref{def:structural_conversation_network}. Generally, interpersonal communication modelled by the structural conversation network implies directionality and a post is usually meant as a response to another, e.g., reply $\rightarrow$ seed post. Therefore, the structural conversation network is a \emph{directed} network. Also, the outgoing degree of a node is one except for the root node, which is not a response to any other post. Therefore, if one would not consider the edges' directionality, two posts would be connected by exactly one path, and the structural conversation network could be described as a \emph{tree} (cf. Cogan et al. \cite{cogan2012reconstruction}).

\subsection{Temporal conversation model}\label{sec:temporal_conversation_model}

Conversations are inherently dynamic and evolve over time. It is crucial to consider these dynamics for modelling a conversation. Typically the posts contained in a conversation come with a timestamp, such as the publishing date. Formally, the assignment of timestamps is described by a node time mapping $\tau: V \rightarrow T$ with $T$ representing the domain of time. Given that the posts' timestamps impose a natural order on a conversation's elements, the respective structural conversation network can be split into a temporal sequence of snapshots.

\begin{definition}[Temporal conversation model]
    Each post in a structural conversation network comes with a timestamp: $\tau(v) = t$. Given a sequence of $n$ timestamps $(t_i)_{i=0}^{n}$ with $t_i < t_{i+1}$, the structural network $G_{struc}$ can be split into $n$ temporal snapshots: $G_{temp} = (G_i)_{i=1}^{n}$. For each snapshot $G_i=(V_i, E_i) \in G_{temp}$ the contained nodes and edges are defined as follows: $V_i = \{v \in V\;|\;t_0 \leq \tau(v) = t < t_i\}$ and $E_i = \{e = (v_i, v_j) \in E\;|\;v_i, v_j \in V_i\}$.
\end{definition}

Notably, the time windows used for the snapshots do not have to be of equal size. The condition $|t_{i+1} - t_i| = |t_{i+3} - t_{i+2}|$ does not necessarily hold. The network $G_{struc}$ might also be sampled according to the volume of conversation posts as an alternative to equally sized time windows. In such a case, the condition $|V_i'| = |\{v \in V\;|\;t_{i-1} \leq \tau(v) = t < t_i\}| = |V_{i+1}'| = |\{v \in V\;|\;t_i \leq \tau(v) = t < t_{i+1}\}|$ holds, meaning that the amount of newly added posts is constant across the network snapshots. This \emph{volume-based sampling} approach is especially suitable in cases where conversation posts are not equally distributed over time. Therefore, a time-based sampling, as suggested by Cogan et al. \cite{cogan2012reconstruction} and used by Mathew et al. \cite{mathew2016mining}, would lead to a highly skewed distribution of newly added posts across the network snapshots.

\subsection{Content-aware conversation model}\label{sec:content_aware_conversation_model}

\begin{figure}[b]
    \centering
    \includegraphics[width=0.6\columnwidth]{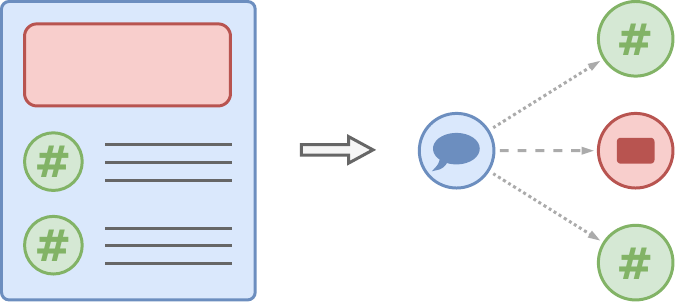}
    \caption{Modelled content objects are extracted from the analyzed posts. Extracted content, the two hashtags and the image, is linked to the respective post node in the content-aware conversation network.}
    \label{fig:extraction_process}
\end{figure}

Next to structure and dynamics, the actual content of a conversation must be considered by a conversation model as well. Various types of semantically meaningful content can be part of a conversation, such as written text, shared URLs, posted images or involved users. These content objects are extracted from the analyzed posts to be incorporated into the conversation network. Figure \ref{fig:extraction_process} illustrates this extraction process. In the content-aware conversation network, the extracted content objects are linked to their belonging post node.

\begin{definition}[Content-aware conversation network]
    The content-aware conversation model is based on the network $G_{struc}$, enriched by additional content objects modelled as nodes. Additional node types $a \in A$ indicate the different content types. The actual content objects are extracted from the respective conversation posts: $\epsilon: V_P \rightarrow 2^{V_C}$ with $V_P = \{v \in V\;|\;\phi(v) = \text{post}\}$ and $V_C=\{v \in V\;|\;\phi(v) \in A\backslash\{\text{post}\}\}$. To indicate the extraction process, in the content-aware conversation network $G_{cont}$ the content objects are linked to the original post node(s): $v_p \rightarrow v_c$ with $v_c \in V_C$ and $v_c \in \epsilon(v_p)$. These links might as well be typed. Respective types are modelled as additional edge types $r \in R$.
\end{definition}

The same content object might belong to different posts in the content-aware conversation network (e.g., the same URL might be shared in multiple posts). Therefore, two formerly disconnected post nodes might become indirectly linked via shared content nodes (e.g., URL, hashtag, username).

The content enrichment of the structural conversation network also applies to the individual snapshots of the temporal conversation network $G_{temp}$. Each network snapshot then contains the content objects belonging to the contained posts. Together, the temporal and content-aware conversation model incorporates a conversation's structure, dynamics and content. Formally, the CODY model can be seen as a THIN.

\subsection{Temporal Wiener index}\label{sec:temporal_wiener_index}

To quantitatively describe the structure of conversation networks, tailored metrics are needed. In a recent work, the Wiener index, known from the field of chemistry \cite{wiener1947structural}, has been leveraged to measure the structure of reply trees \cite{saveski2021structure}. According to Saveski et al. \cite{saveski2021structure}, the Wiener index describes two characteristic topologies of a tree-like graph in its extreme cases. On the one hand, a low Wiener index indicates a conversation in which all non-seed posts respond to the original seed post, and these non-seed posts do not show any interactions among each other. On the other hand, a high Wiener index indicates a conversation tree with a single dominant interaction branch. To the best of our knowledge, no work extends the Wiener index to fit temporally evolving conversation trees. Based on the definition of the static Wiener index employed by Saveski et al. \cite{saveski2021structure}, we define the \emph{temporal} Wiener index as formalized in the following:

\begin{definition}[Temporal Wiener index]
    The temporal Wiener index $w(G_{temp})$ of a temporal sequence $(G_i)_{i=1}^{n}$ of tree-like structured conversation networks is determined by the average distance between post nodes in each network snapshot:

    \begin{equation}
        w(G_{temp}) = \left(\frac{1}{|V_i|\cdot(|V_i|-1)}\sum_{m \in V_i}\sum_{n \in V_i} \delta_i(m, n)\right)_{i=1}^{n}
    \end{equation}

    with $\delta_i(m, n)$ denoting the length of the shortest path between the nodes $m$ and $n$ in the temporal snapshot $G_i$.
\end{definition}

Given a temporal sequence of conversation networks, the temporal Wiener index thus itself is given as a temporal sequence of static Wiener indices. Therefore, the time series of Wiener indices allows for studying the temporal changes of the Wiener index across the temporal evolution of a conversation.
\section{Experiments}\label{sec:experiments}

To demonstrate the capabilities of the CODY model, this section discusses the results of various conducted experiments. For all experiments, a Twitter conversation dataset that we collected is leveraged (cf. Section \ref{sec:dataset}). Also, our developed software package (cf. Section \ref{sec:introduction}) is used to analyze the experiments' results with regard to HIN analysis.

\subsection{Twitter conversation dataset}\label{sec:dataset}

\begin{figure}[t]
    \centering
    \includegraphics[width=\columnwidth]{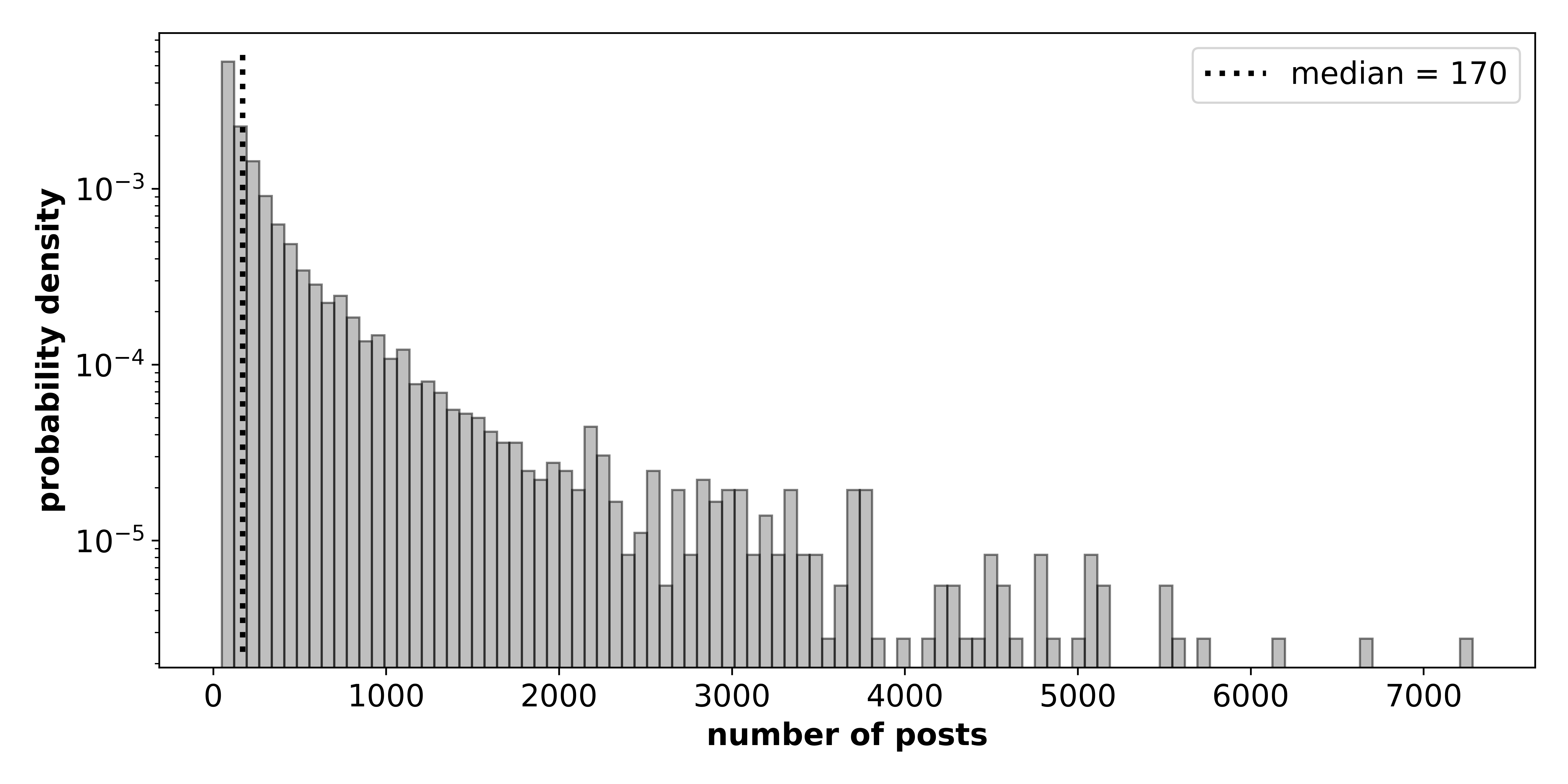}
    \caption{Distribution of conversation sizes as contained in the leveraged dataset of the German political Twittersphere.}
    \label{fig:conversation_sizes}
\end{figure}

The dataset used for the subsequent experiments is based on a collection of German political Twitter accounts published as EPINetz Twitter dataset \cite{konig2022epinetz}. We collected conversations that started from a seed post of one of these accounts by leveraging the official Twitter search API v2\footnote{Twitter Developer Platform: \url{https://developer.twitter.com/en/docs/twitter-api/tweets/search/introduction}, accessed 10/07/23}. According to Maireder and Ausserhofer, the political domain is particularly suitable for examining the structure of conversations (\cite{maireder2014political}, as cited in \cite{koylu2019modeling}): "Retweets and mention/reply functions have been found to influence the structure of conversational discourse if it is especially related to politics." In total, the dataset consists of more than 2M tweets as part of 5k conversations, with each conversation including at least 50 posts. Figure \ref{fig:conversation_sizes} shows the distribution of conversation sizes. The tweets were published between December 2020 and January 2023, and the median conversation duration in the dataset is about seven days.

\begin{figure}[t]
    \centering
    \includegraphics[width=0.5\columnwidth]{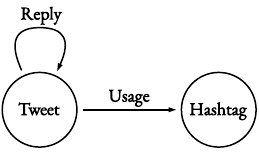}
    \caption{Network schema used for the conducted experiments. A conversation's reply tree consist of tweet nodes connected by replies. Further, the "usage" relationship indicates that a hashtag is used within a tweet.}
    \label{fig:network_schema}
\end{figure}

According to Sun and Han, the Twitter information network is well suited to be modelled as HIN \cite{sun2013mining}. In line with the CODY model, we view each conversation as a temporal and content-aware conversation network with the contained tweets as posts and hashtags as content objects. Figure \ref{fig:network_schema} illustrates the network schema of the conversation model used for the subsequent experiments. In the analyzed networks, links between tweets indicate replies which are seen as "[\ldots] the most direct communication sign between two users" \cite{cogan2012reconstruction} and therefore, make up the structure of a conversation.

Regarding data quality, there is no guarantee that the complete lifetime of a conversation is captured. This would require a continued re-crawl of the conversation data, which is unfeasible due to API usage restrictions. Nevertheless, the reported high average conversation duration suggests that most of the conversation's evolution is captured. This assumption is also underlined by the conducted experiment, which examines the posting activity of the collected conversations (cf. Section \ref{sec:posting_activity}). Most activities seem to occur within the first 10\% of a conversation's lifetime. Therefore, sufficient posts should be contained in the dataset even for shorter conversations, for which data of less than a day is captured. In their analysis, Cogan et al. state that conversation activity usually stops after 6 hours \cite{cogan2012reconstruction}. Further, additional data quality issues related to online conversations have been reported in the past, such as replies to private profiles or the deletion of posts \cite{cogan2012reconstruction}. However, less than 5\% of conversations seem to be affected by that \cite{cogan2012reconstruction}. Missing posts lead to a split of the conversation graph into multiple components. Therefore, we take the weakly connected, giant component for creating the reply trees to handle the cases in which already deleted posts or those that link to private profiles are part of a conversation. Also, in a prior step to generate these reply trees, tweets are grouped into conversations based on the \texttt{conversation\_id} field contained in the raw API payload. Among the tweets of a conversation, reply links are determined by the additional \texttt{referenced\_tweets} field and the contained entries of type \texttt{replied\_to}.

Even though the following experiments have been conducted based on the described dataset, we argue that the CODY framework is applicable to a wide range of use cases and conversation data of platforms different from Twitter could be used as well. The primary assumption of the model, which is to have interrelated and timestamped conversation posts, is fulfilled by a large variety of online conversation data sources.

\subsection{Posting activity}\label{sec:posting_activity}

\begin{figure}[t]
    \centering
    \includegraphics[width=\columnwidth]{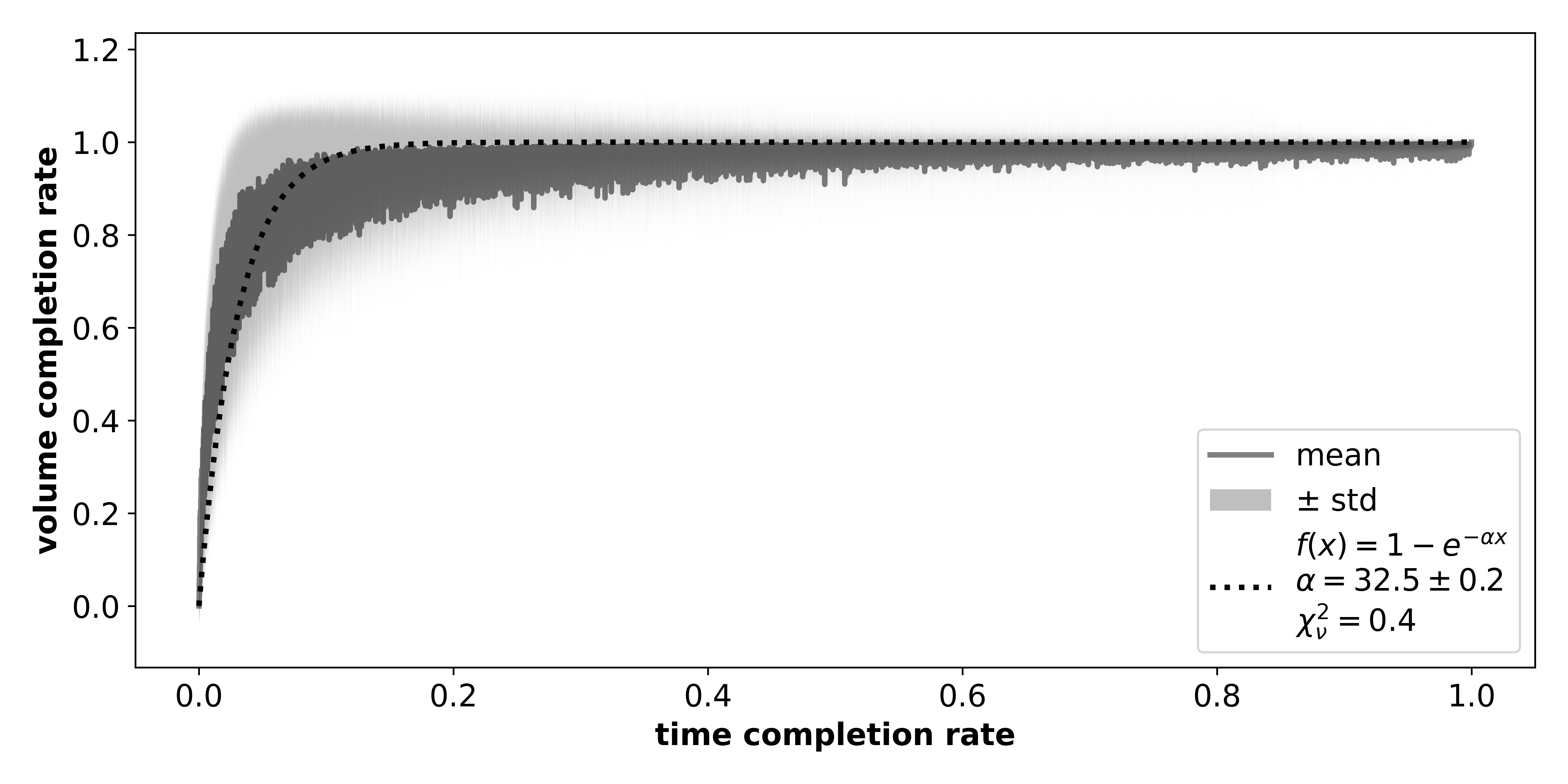}
    \caption{The evolution of a conversation regarding the number of posts is characterized by an exponential saturation function.}
    \label{fig:completion_fit}
\end{figure}

Analyzing the posting activity across the lifetime of a conversation is crucial for a more general understanding of conversation dynamics (cf. \ref{item:rq1}). It lays the foundation for deciding which temporal sampling approach must be leveraged to model the conversation dynamics correctly (cf. Section \ref{sec:temporal_conversation_model}). Figure \ref{fig:completion_fit} shows the volume completion rate (relative number of posts) of the conversations in our dataset plotted against the conversations' lifetime (time completion rate). The dark and lighter value bands indicate mean values and their standard deviations. Values are calculated using a $10^{-5}$ resolution of the completion rate. The large fluctuations show that temporal posting activity varies a lot between conversations. Still, a general activity pattern across the lifetime of a conversation can be observed. If $\lambda_{volume}$ denotes the volume completion rate of a conversation, which can also be described as the fraction of the total amount of conversation posts, and $\lambda_{time}$ denotes the time completion rate (lifetime) of a conversation, a conversation's posting activity can be modelled as

\begin{equation}\label{ref:posting_activity}
    \lambda_{volume} = 1 - e^{-\alpha\lambda_{time}}
\end{equation}

with $\alpha$ as the saturation rate. As shown by the fitted function in Figure \ref{fig:completion_fit} and indicated by a reduced chi-square value, $\chi^2_{\nu}$, of $0.4$, this model, despite its simplicity, describes the process in a precise manner. For the curve fitting, the \texttt{lmfit} Python package is used \cite{newville2016lmfit}. Most of the posting activity happens during the early stage of a conversation. After an exponential increase in conversation posts, the posting activity slows down and converges towards its final saturation level. This saturating process is also characterized by the saturation time constant $\Gamma$, defined as $\Gamma = \frac{1}{\alpha}$. For our dataset, the saturation time is about $0.03$, meaning that on average, after 3\% of the conversation lifetime, about $1-\frac{1}{e} \approx 0.63\;\widehat{=}\;63\%$ of the posts are published.

The non-linear activity pattern must be considered when selecting the sampling method for the temporal modelling of conversations (cf. Section \ref{sec:temporal_conversation_model}). Given the exponential increase in posts during the early stage of a conversation, a sampling approach using fixed time windows would lead to a highly skewed distribution of newly added posts across the network snapshots of a conversation. In the beginning, the network snapshots would grow significantly, whereas in later stages, only minor changes would be captured. Quantitatively the change rate of sampling snapshots can be described by the derivative of the posting activity:

\begin{equation}\label{ref:snapshot_change_rate}
    \lambda_{volume}' = \frac{d}{d\lambda_{time}}\lambda_{volume} = \alpha e^{-\alpha\lambda_{time}}.
\end{equation}

Therefore, the ratio of snapshot change rates during two points in time $\lambda_{time_1}$ and $\lambda_{time_2}$ of a conversation's lifetime is given as:

\begin{equation}\label{ref:change_rate}
    \frac{\lambda_{volume}'(\lambda_{time_1})}{\lambda_{volume}'(\lambda_{time_2})} = e^{\alpha(\lambda_{time_2}-\lambda_{time_1})}.
\end{equation}

For our conversation activity model ($\alpha = 32.5$), about $131$ times more posts would be added to a snapshot right at the beginning of a conversation ($\lambda_{time_1}=0.05$) compared to a later stage ($\lambda_{time_2} = 0.2$) by using fixed time windows. Therefore, to correctly capture changes across the entire lifetime of a conversation, the exponential saturation regarding the posting activity needs to be considered. This can easily be done using fixed volume sampling instead of fixed time windows. The conversation size difference between two successive snapshots is kept constant for this. In our case, we determine a sampling rate of five new posts per snapshot. This sampling approach is also used for the subsequent experiments. Given that the relative volume of posted tweets can be seen as \emph{normalized} conversation time, we use a conversation's completion rate as the unit for the temporal dimension whenever appropriate.

\subsection{Temporal Wiener index}\label{sec:wiener_index_experiment}

\begin{figure}[t]
    \centering
    \includegraphics[width=\columnwidth]{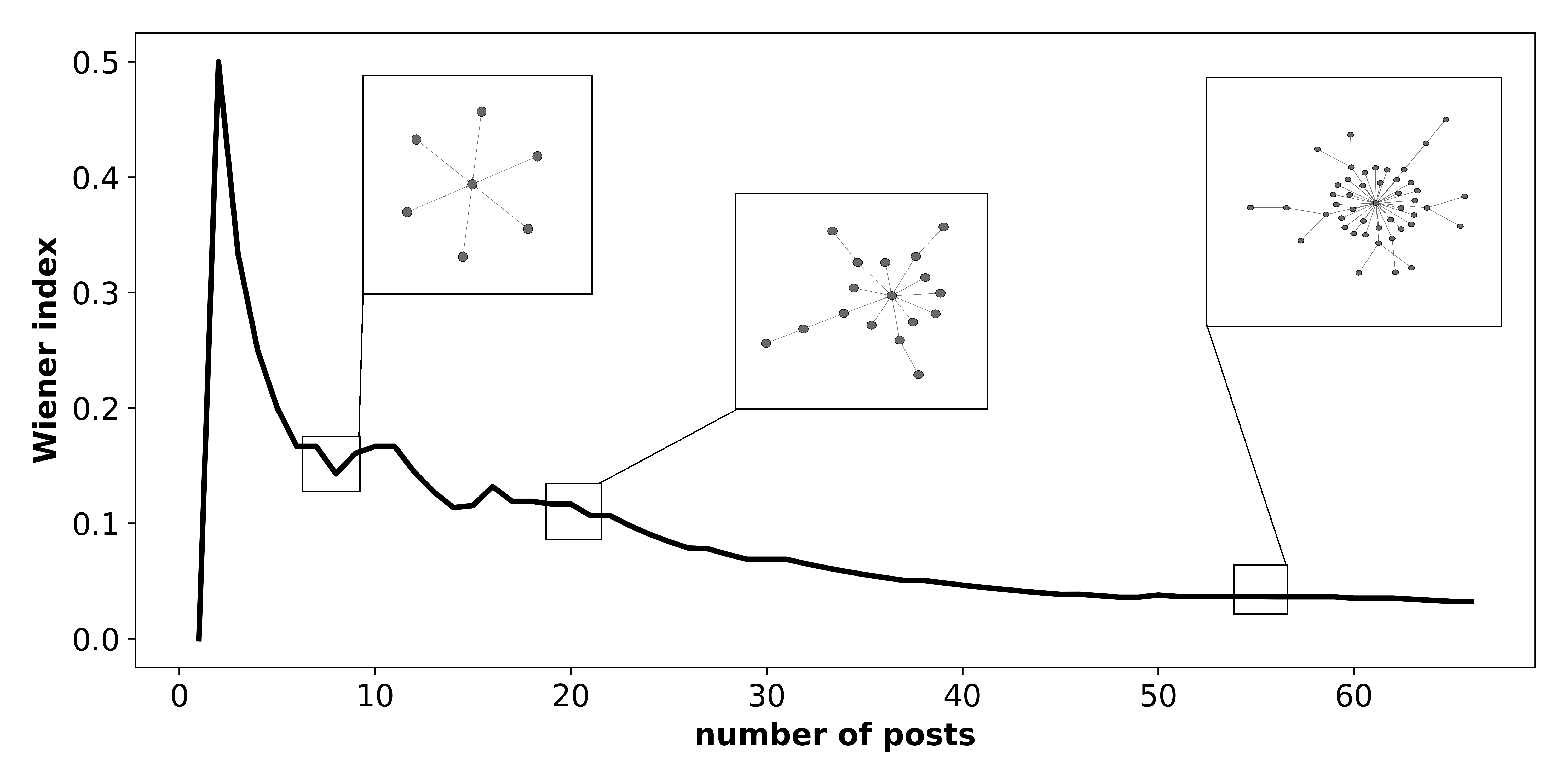}
    \caption{Exemplary temporal Wiener index. After a steep increase at the beginning of the conversation, the value rapidly decreases. Structural changes in the conversation also reflect this evolution. Over time most posts directly link to the original seed post, and offshoots are rarely observed.}
    \label{fig:wiener_index_example}
\end{figure}

The CODY framework views conversations as networks. This approach allows us to analyze the conversations' topological structure. Leveraging the \emph{temporal} Wiener index, this can also be done by considering the conversation's dynamics (cf. \ref{item:rq2}). For the following analysis, we calculate the temporal Wiener indices for the conversations contained in our dataset. Only their reply trees are considered without additional content objects (i.e., hashtags) to comply with the definition of the Wiener index metric. For the software implementation, the \texttt{igraph} Python package is used \cite{csardi2006igraph}. To handle the edge case of a single-node reply tree, we define the Wiener index as zero at this conversation state.

Figure \ref{fig:wiener_index_example} shows a temporal Wiener index example. As one can observe, the structural evolution of the network is also reflected by the temporal change of the Wiener index. The initial rise of the Wiener index value is caused by the first reply, which directly links to the original seed post. Given that only one shortest path of length one is contained in the network at this stage, namely from the first reply post to the original post, a maximum Wiener index value of $0.5$ is derived for this network snapshot. From there on, most posts are directly linked to the seed post, reflected by the rapid decrease of the temporal Wiener index. Overall, off-shooting branches are rarely observed in the conversation network, which explains its low final value of about $0.03$.

\begin{figure}[t]
    \centering
    \includegraphics[width=\columnwidth]{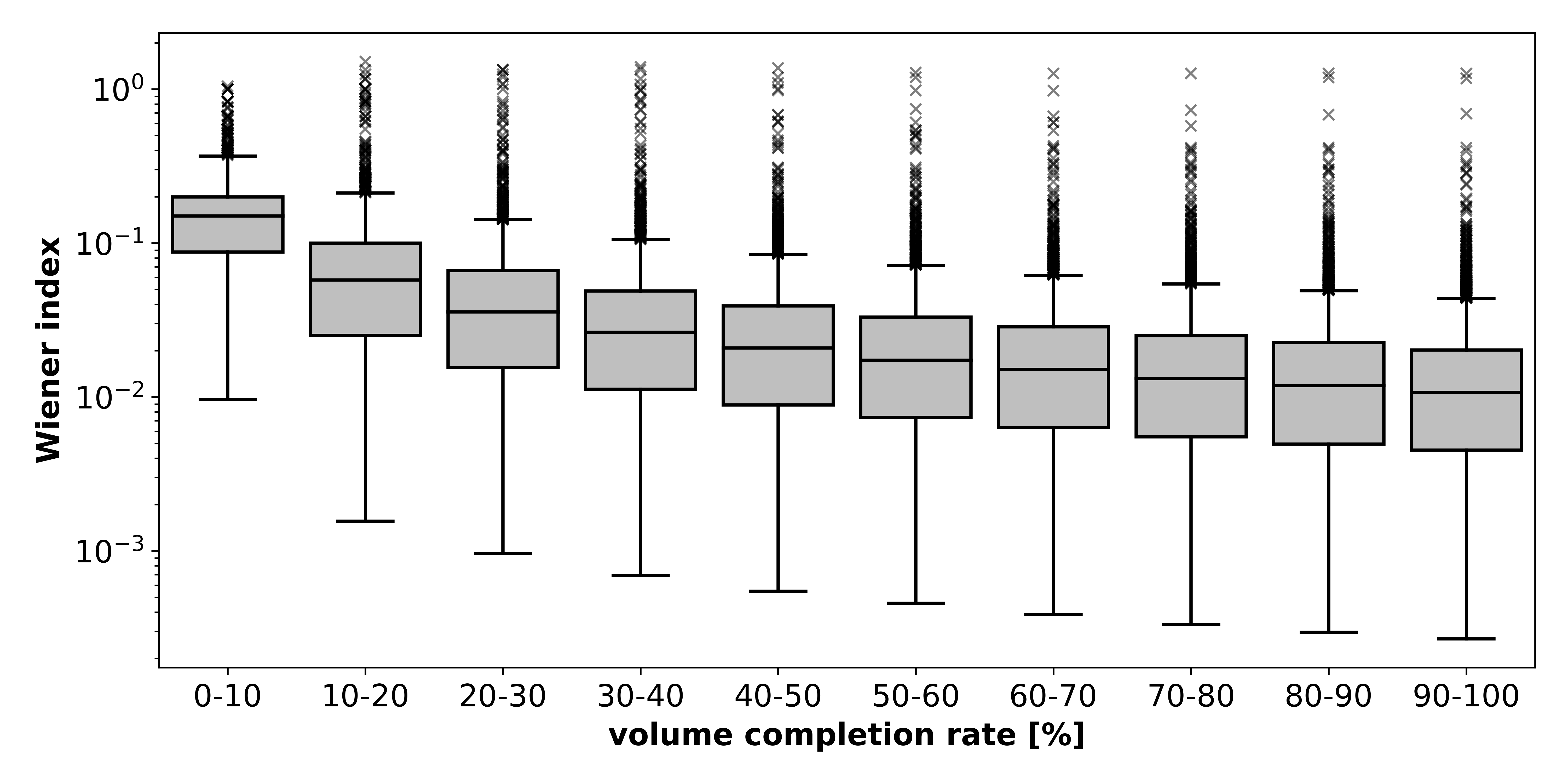}
    \caption{Box plot showing the averaged temporal evolution of Wiener index values across the lifetime of the analyzed conversations. Decreasing values indicate the tendency of conversations to become less sprawling.}
    \label{fig:wiener_index_evolution}
\end{figure}

To not just focus on a single conversation example but to analyze the structural evolution of conversations in a more general way, the average temporal Wiener index across the lifetime of a conversation is shown in Figure \ref{fig:wiener_index_evolution} in the form of a box plot. First, it should be noted that temporal Wiener index values vary a lot between the different conversations, as shown by the large quartile ranges, the high variability, and the outliers. Still, a general evolution pattern can be observed. Over time, the Wiener index decreases and converges towards a final median value of about $0.01$. Structurally, this means that over time, most replies directly link to the original seed post, and sprawling conversation branches do not seem to play a significant role in the conversation network. Nevertheless, to return to the high data variability, in rare cases, offshoots might be observed in the conversation structure, reflected by the higher outlier values as shown in the box plot. Interestingly, Cogan et al. report that more than 60\% of their analyzed Twitter conversation graphs follow the structure of paths \cite{cogan2012reconstruction}. Apart from the described outliers, this structure is less prevalent in our case as it would lead to significantly higher final Wiener index values. Many reasons are conceivable for this structural difference. One might be that, in our case, most conversation participants wanted to interact directly with the political actor who posted the original seed post. Thus, more star-like conversation structures are observed.

\subsection{Hashtag Hijacking}\label{sec:hashtag_hijacking}

In a holistic conversation analysis framework, the discussed content and its dynamics should also be considered next to the structure of the conversation. On the Twitter platform, used hashtags are an indicator of the discussed topics and the semantics of posted content. In that sense, hashtags can be leveraged to analyze how the topical focus of a conversation changes over time. More specifically, one can investigate whether initially used hashtags keep their dominant prevalence over the lifetime of a conversation (cf. \ref{item:rq3}). For this, we leverage the proposed Twitter conversation dataset and, given that hashtags are represented as nodes in the content-aware conversation network, temporally track their degree centrality as an indicator of the importance they play in the conversation. In the second step, the temporal degree evolution of the hashtag(s) used in the initial seed post is compared to the evolution of the other hashtags used later in the conversation. Suppose a non-initial hashtag reaches a higher degree centrality value than the initially used hashtag(s) at the end of a conversation; we denote the scenario as \emph{hashtag hijacking}. In case of an overtake that is followed by a "retake" of an initial hashtag so that it reclaims the highest prevalence with regards to its degree centrality in the conversation network at the conversation's ending, the scenario is called \emph{failed hijacking}. It is not a hijacking scenario if no non-initial hashtags are used in a conversation, or no overtake happens. If multiple hashtags take over the initial hashtag(s), the scenario is only counted as a single hijacking. Semantically, a hashtag hijacking scenario can be seen as an indicator of a \emph{topical shift} within a conversation. Nevertheless, it has to be noted that at this point, we do not employ any natural language processing techniques to analyze the semantic similarity of the prevalent hashtags in a conversation and whether these are used synonymously. Takeovers might as well be performed by semantically similar hashtags, and therefore, only a slight semantic shift would be caused. An extension to overcome this shortcoming would be a valuable future contribution.

To apply the methodology, we focus on conversations where the initial seed post contains at least one hashtag. Also, conversations should have at least one hashtag that is used a sufficient number of times. In our case, we set this threshold to five usages in a conversation. About 1.6k of the 5k conversations in the analyzed Twitter corpus fulfil the outlined requirements. In total, about 1k hijacking scenarios ($\sim 67\%$) are detected in these conversations. Further, 111 failed hijacking scenarios ($\sim 7\%$) and 427 scenarios without hijacking ($\sim 26\%$) are found. Compared to the total number of conversations in the dataset, successful hashtag hijackings comprise a considerable portion of about $22\%$.

\begin{figure}[t]
    \centering
    \includegraphics[width=\columnwidth]{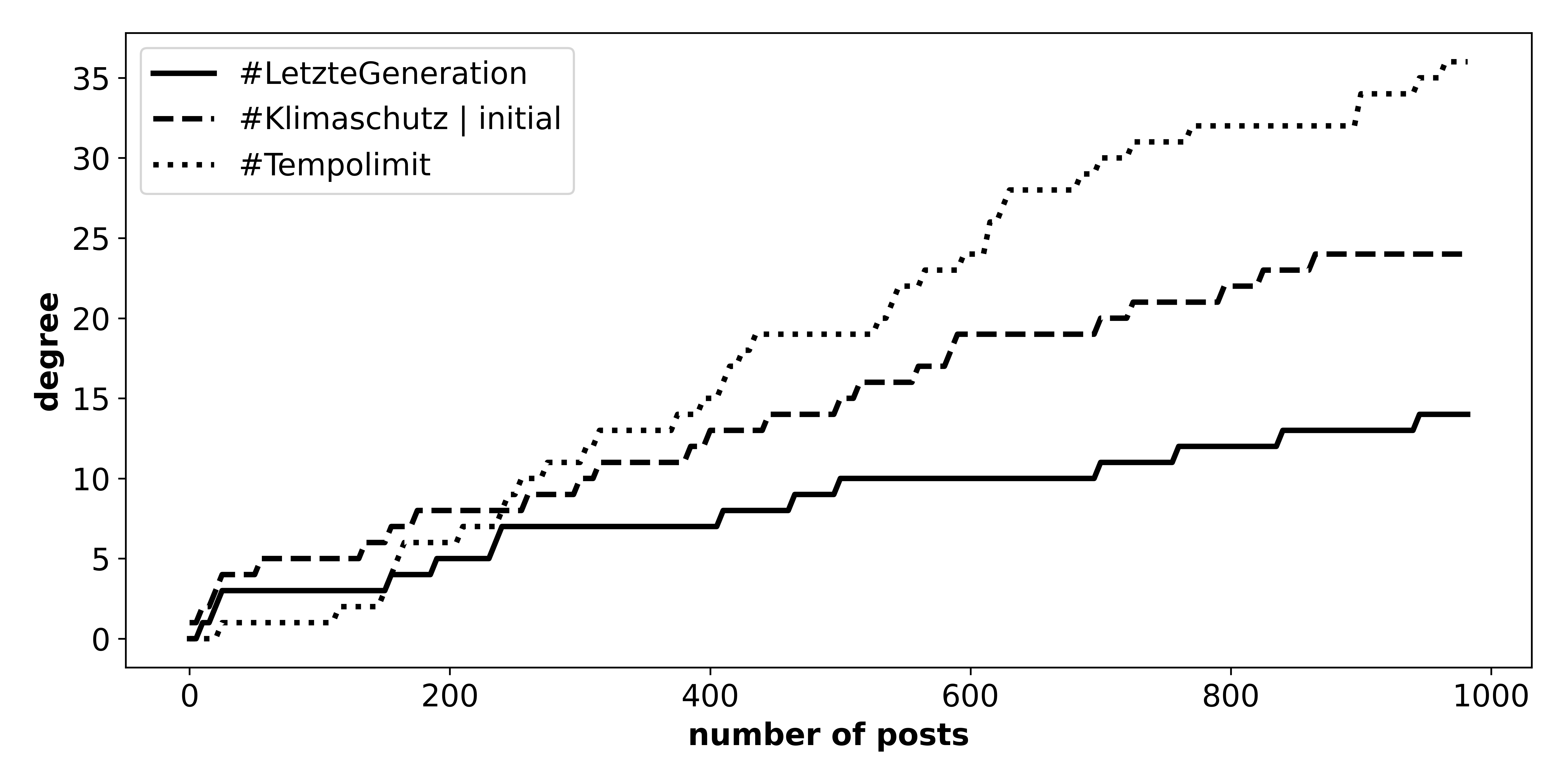}
    \caption{Example of a hashtag hijacking scenario. The initial hashtag \emph{Klimaschutz} (Eng. climate protection) is overtaken by the hashtag \emph{Tempolimit} (Eng. speed limit) in terms of degree centrality.}
    \label{fig:hijacking_example}
\end{figure}

As an example of a hijacking scenario, Figure \ref{fig:hijacking_example} shows the temporal degree centrality of three hashtags used in the related Twitter conversation\footnote{Twitter conversation: \url{https://twitter.com/Wissing/status/1596101821994606592}, accessed 10/07/23}. The original tweet of the conversation criticizes a climate protest by the "Letzte Generation" (Eng. "Last Generation") movement, in which activists glued themselves to the tarmac of an airport in order to interrupt flight operations. The \emph{Klimaschutz} (Eng. climate protection) hashtag is used in this tweet. Interestingly, the initial thematic positioning of the tweet is overtaken. This takeover is reflected by the hashtag hijacking of the \emph{Tempolimit} (Eng. speed limit) hashtag, which represents the discussion about a speed limit on German autobahns. In this case, the hijacking can be understood as a counter-positioning of many conversation participants. Alternatively, their intention might have been to relate this conversation to another climate-related conversation.

\begin{figure}[t]
    \centering
    \includegraphics[width=\columnwidth]{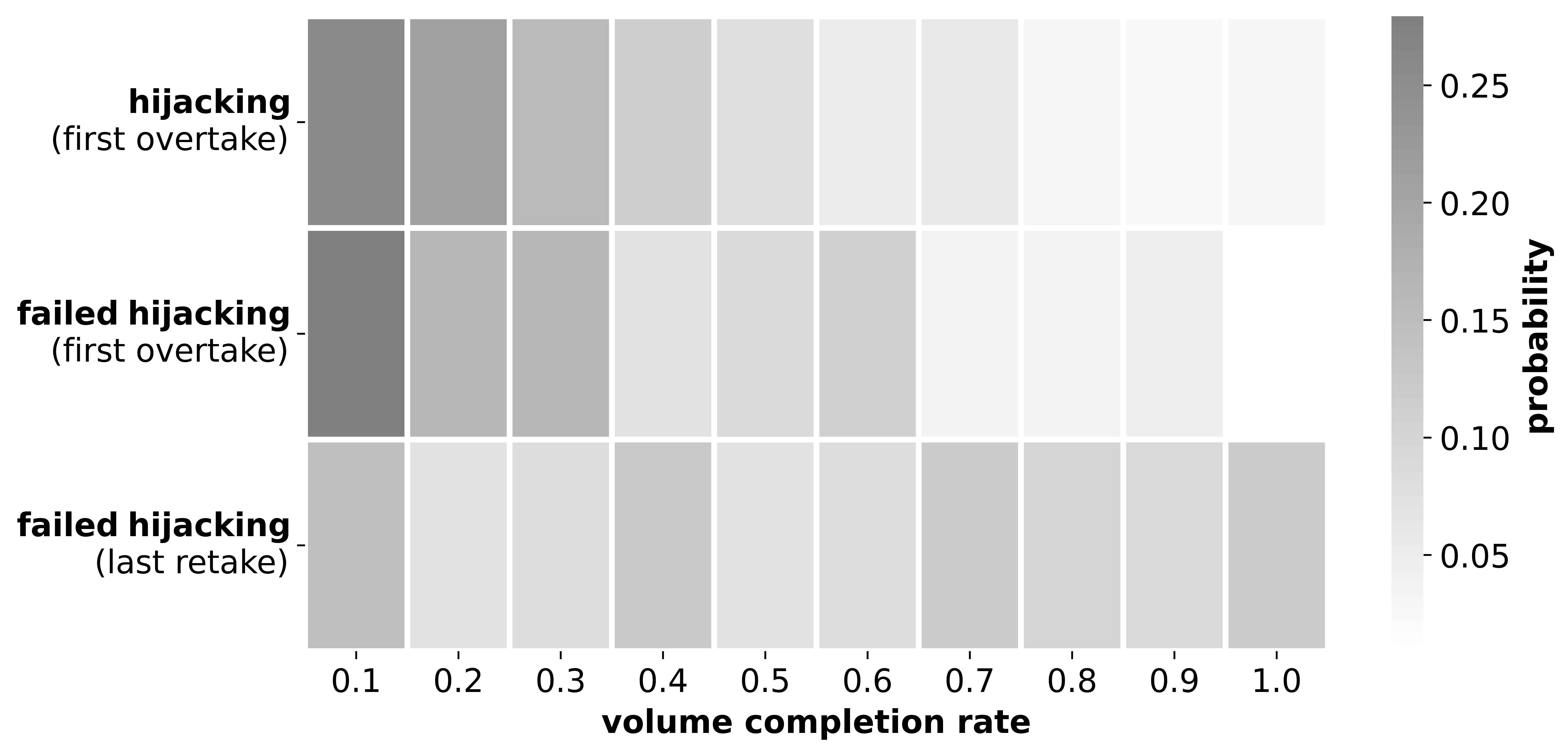}
    \caption{Initial hashtag overtakes in hijacking scenarios mainly happen at the beginning of a conversation. Unlike in successful scenarios, in failed ones and for retakes, the overtake tends to occur slightly later in the conversation.}
    \label{fig:hijacking_heatmap}
\end{figure}

Apart from a hashtag hijacking scenario occurring within a conversation, some use cases might also benefit from an analysis of \emph{when} the takeovers happen. Figure \ref{fig:hijacking_heatmap} shows a heatmap of the occurrence probability of the takeovers related to the (failed) hijacking scenarios across the lifetime of a conversation. Dark patches indicate a high takeover occurrence probability, whereas bright ones indicate the opposite. All heatmap rows are normalized based on the total number of occurrences of the respective takeover. As timestamps of the occurrences, we take, on the one hand, the first point at which a takeover is happening by the finally dominating hashtag. For both scenarios, this is when a non-initial hashtag takes over the initial hashtag(s). On the other hand, for the failed hijacking scenario, we also keep track of the last time a "retake" of one of the initial hashtags occurs. At this point, one of the initial hashtags is reclaiming its position as the most prevalent hashtag in the conversation. As shown by Figure \ref{fig:hijacking_heatmap}, many initial takeovers happen early in the conversations. A slight shift towards the end of the conversation can be observed for the failed hijacking scenario. Nevertheless, the "competition" for the most prevalent position is not finished at this point. Towards the end of the conversations, most retakes occur.

\section{Conclusion and Future Work}\label{sec:conclusion}

Existing work that focuses on analyzing conversations in online social networks only partially considers the three dimensions of a conversation: structure, content and dynamics. To address this shortcoming, the CODY model is presented in this work. Based on the concept of temporal heterogeneous information networks, it allows to model a conversation's topological structure, incorporate various content objects, and consider its temporal evolution. To demonstrate the model's capabilities, results of several experiments are presented. These experiments cover the examination of posting activity across the lifetime of a conversation, the study of our novel temporal Wiener index metric, and the analysis of various "hashtag hijacking" scenarios. Further work will employ more advanced natural language processing methods to improve the semantic understanding of analyzed conversations. Also, framing the hashtag hijacking phenomenon as a prediction task might be an interesting future extension. In this regard, one might also examine the relationship between a conversation's structural and content-wise evolution.

\bibliographystyle{IEEEtran}
\bibliography{IEEEabrv,references}

\begin{thebibliography}{10}
\providecommand{\url}[1]{#1}
\csname url@samestyle\endcsname
\providecommand{\newblock}{\relax}
\providecommand{\bibinfo}[2]{#2}
\providecommand{\BIBentrySTDinterwordspacing}{\spaceskip=0pt\relax}
\providecommand{\BIBentryALTinterwordstretchfactor}{4}
\providecommand{\BIBentryALTinterwordspacing}{\spaceskip=\fontdimen2\font plus
\BIBentryALTinterwordstretchfactor\fontdimen3\font minus \fontdimen4\font\relax}
\providecommand{\BIBforeignlanguage}[2]{{%
\expandafter\ifx\csname l@#1\endcsname\relax
\typeout{** WARNING: IEEEtran.bst: No hyphenation pattern has been}%
\typeout{** loaded for the language `#1'. Using the pattern for}%
\typeout{** the default language instead.}%
\else
\language=\csname l@#1\endcsname
\fi
#2}}
\providecommand{\BIBdecl}{\relax}
\BIBdecl

\bibitem{brambilla2022graph}
M.~Brambilla, A.~Javadian~Sabet, K.~Kharmale, and A.~E. Sulistiawati, ``{Graph-Based Conversation Analysis in Social Media},'' \emph{Big Data and Cognitive Computing}, vol.~6, no.~4, p. 113, 2022.

\bibitem{saveski2021structure}
M.~Saveski, B.~Roy, and D.~Roy, ``{The structure of toxic conversations on Twitter},'' in \emph{Proc. of the Web Conference 2021}, 2021, pp. 1086--1097.

\bibitem{cogan2012reconstruction}
P.~Cogan, M.~Andrews, M.~Bradonjic, W.~S. Kennedy, A.~Sala, and G.~Tucci, ``{Reconstruction and analysis of Twitter conversation graphs},'' in \emph{Proc. of the First ACM International Workshop on Hot Topics on Interdisciplinary Social Networks Research}, 2012, pp. 25--31.

\bibitem{sun2010community}
Y.~Sun, J.~Tang, J.~Han, M.~Gupta, and B.~Zhao, ``{Community evolution detection in dynamic heterogeneous information networks},'' in \emph{Proc. of the Eighth Workshop on Mining and Learning with Graphs}, 2010, pp. 137--146.

\bibitem{li2018temporal}
Y.~Li, Z.~Lou, Y.~Shi, and J.~Han, ``{Temporal motifs in heterogeneous information networks},'' in \emph{MLG Workshop@KDD}, 2018.

\bibitem{milani2019relationship}
A.~Milani~Fard, E.~Bagheri, and K.~Wang, ``{Relationship Prediction in Dynamic Heterogeneous Information Networks},'' in \emph{Advances in Information Retrieval}.\hskip 1em plus 0.5em minus 0.4em\relax Cham: Springer, 2019, pp. 19--34.

\bibitem{botzer2023entity}
N.~Botzer and T.~Weninger, ``Entity graphs for exploring online discourse,'' \emph{Knowledge and Information Systems}, pp. 1--19, 2023.

\bibitem{mathew2016mining}
B.~Mathew, T.~Chakraborty, N.~Ganguly, and S.~Datta, ``{Mining Twitter conversations around E-commerce promotional events},'' in \emph{Proc. of the 19th ACM Conference on Computer Supported Cooperative Work and Social Computing Companion}, 2016, pp. 345--348.

\bibitem{guille2013information}
A.~Guille, H.~Hacid, C.~Favre, and D.~A. Zighed, ``Information diffusion in online social networks: A survey,'' \emph{ACM Sigmod Record}, vol.~42, no.~2, pp. 17--28, 2013.

\bibitem{yujie2020survey}
Y.~Yujie, ``{A survey on information diffusion in online social networks},'' in \emph{Proc. of the 2020 European Symposium on Software Engineering}, 2020, pp. 181--186.

\bibitem{cuzzocrea2013community}
A.~Cuzzocrea and F.~Folino, ``{Community evolution detection in time-evolving information networks},'' in \emph{Proc. of the Joint EDBT/ICDT 2013 Workshops}, 2013, pp. 93--96.

\bibitem{sajadmanesh2019continuous}
S.~Sajadmanesh, S.~Bazargani, J.~Zhang, and H.~R. Rabiee, ``{Continuous-time relationship prediction in dynamic heterogeneous information networks},'' \emph{ACM Transactions on Knowledge Discovery from Data (TKDD)}, vol.~13, no.~4, pp. 1--31, 2019.

\bibitem{wang2020dynamic}
X.~Wang, Y.~Lu, C.~Shi, R.~Wang, P.~Cui, and S.~Mou, ``{Dynamic heterogeneous information network embedding with meta-path based proximity},'' \emph{IEEE Transactions on Knowledge and Data Engineering}, vol.~34, no.~3, pp. 1117--1132, 2020.

\bibitem{bian2019network}
R.~Bian, Y.~S. Koh, G.~Dobbie, and A.~Divoli, ``{Network embedding and change modeling in dynamic heterogeneous networks},'' in \emph{Proc. of the 42nd International ACM SIGIR Conference on Research and Development in Information Retrieval}, 2019, pp. 861--864.

\bibitem{xie2021sequential}
T.~Xie, Y.~Xu, L.~Chen, Y.~Liu, and Z.~Zheng, ``{Sequential recommendation on dynamic heterogeneous information network},'' in \emph{2021 IEEE 37th International Conference on Data Engineering (ICDE)}.\hskip 1em plus 0.5em minus 0.4em\relax IEEE, 2021, pp. 2105--2110.

\bibitem{jia2017link}
Y.~Jia, Y.~Wang, X.~Jin, Z.~Zhao, and X.~Cheng, ``{Link inference in dynamic heterogeneous information network: A knapsack-based approach},'' \emph{IEEE Transactions on Computational Social Systems}, vol.~4, no.~3, pp. 80--92, 2017.

\bibitem{aggarwal2012dynamic}
C.~Aggarwal, Y.~Xie, and P.~S. Yu, ``{On dynamic link inference in heterogeneous networks},'' in \emph{Proc. of the 2012 SIAM International Conference on Data Mining}.\hskip 1em plus 0.5em minus 0.4em\relax SIAM, 2012, pp. 415--426.

\bibitem{sun2013mining}
Y.~Sun and J.~Han, ``{Mining heterogeneous information networks: a structural analysis approach},'' \emph{ACM SIGKDD Explorations Newsletter}, vol.~14, no.~2, pp. 20--28, 2013.

\bibitem{wiener1947structural}
H.~Wiener, ``{Structural determination of paraffin boiling points},'' \emph{Journal of the American chemical society}, vol.~69, no.~1, pp. 17--20, 1947.

\bibitem{konig2022epinetz}
T.~K{\"o}nig, W.~J. Sch{\"u}nemann, A.~Brand, J.~Freyberg, and M.~Gertz, ``{The EPINetz Twitter Politicians Dataset 2021. A New Resource for the Study of the German Twittersphere and Its Application for the 2021 Federal Elections},'' \emph{Polit. Vierteljahresschrift}, vol.~63, no.~3, pp. 529--547, 2022.

\bibitem{maireder2014political}
A.~Maireder and J.~Ausserhofer, ``{Political discourses on Twitter: Networking topics, objects, and people},'' \emph{Twitter and society}, vol.~89, pp. 305--318, 2014.

\bibitem{koylu2019modeling}
C.~Koylu, ``{Modeling and visualizing semantic and spatio-temporal evolution of topics in interpersonal communication on Twitter},'' \emph{International Journal of Geographical Information Science}, vol.~33, no.~4, pp. 805--832, 2019.

\bibitem{newville2016lmfit}
M.~Newville, T.~Stensitzki, D.~B. Allen, M.~Rawlik, A.~Ingargiola, and A.~Nelson, ``{LMFIT: Non-linear least-square minimization and curve-fitting for Python},'' \emph{Astrophysics Source Code Library}, pp. ascl--1606, 2016.

\bibitem{csardi2006igraph}
G.~Csardi, T.~Nepusz \emph{et~al.}, ``{The igraph software package for complex network research},'' \emph{InterJournal, complex systems}, vol. 1695, no.~5, pp. 1--9, 2006.

\end{thebibliography}

\end{document}